\documentclass{ws-ijmpcs}

\RequirePackage{xspace}
\usepackage{relsize}

\def\pep2{PEP-II \xspace}

\def\babar{\mbox{\slshape B\kern-0.1em{\smaller A}\kern-0.1em
    B\kern-0.1em{\smaller A\kern-0.2em R}}\xspace}

\def\invfb   {\ensuremath{\mbox{\,fb}^{-1}}\xspace}
\def\epem       {\ensuremath{e^+e^-}\xspace}

\def\CP                {\ensuremath{C\!P}\xspace}

\def\qqbar {\ensuremath{q\overline q}\xspace}

\def\Y#1S{\ensuremath{\Upsilon{(#1S)}}\xspace}

\def\Bbar    {\kern 0.18em\overline{\kern -0.18em B}{}\xspace}

\def\BB      {\ensuremath{B\Bbar}\xspace} 

\def\Kbar  {\kern 0.2em\overline{\kern -0.2em K}{}\xspace}

\def\Kz    {\ensuremath{K^0}\xspace}
\def\Kzb   {\ensuremath{\Kbar^0}\xspace}
\def\KzKzb {\ensuremath{\Kz \kern -0.16em \Kzb}\xspace}
\def\KS    {\ensuremath{K^0_{\scriptscriptstyle S}}\xspace}

\def\Dbar    {\kern 0.2em\overline{\kern -0.2em D}{}\xspace}

\def\Dz      {\ensuremath{D^0}\xspace}
\def\Dzb     {\ensuremath{\Dbar^0}\xspace}

\newcommand{\gevc}{\ensuremath{{\mathrm{\,Ge\kern -0.1em V\!/}c}}\xspace}
\newcommand{\mevc}{\ensuremath{{\mathrm{\,Me\kern -0.1em V\!/}c}}\xspace}
\newcommand{\gevcc}{\ensuremath{{\mathrm{\,Ge\kern -0.1em V\!/}c^2}}\xspace}
\newcommand{\mevcc}{\ensuremath{{\mathrm{\,Me\kern -0.1em V\!/}c^2}}\xspace}

\newcommand{\stat}{\ensuremath{\mathrm{(stat)}}\xspace}
\newcommand{\syst}{\ensuremath{\mathrm{(syst)}}\xspace}

\begin{document}

\markboth{Chunhui Chen}
{\CP\ Violation and Mixing in Charm Meson Decays from \babar}

%
\catchline{}{}{}{}{}
%

\title{\CP\ Violation and Mixing in Charm Meson Decays from \babar }

\author{Chunhui Chen}

\address{Department of Physics and Astronomy\\ 
Iowa State University \\
Ames, Iowa 50010, USA\\
cchen23@iastate.edu\\
(representing the BABAR Collaboration)}

\maketitle

\begin{history}
\received{Day Month Year}
\revised{Day Month Year}
\end{history}

\begin{abstract}
Mixing and \CP\ violation in charm meson decays provide a unique probe of
possible physics beyond the standard model. In this paper, we give a brief
review of the current measurements from the \babar\ experiment.

\end{abstract}

\ccode{PACS numbers: 11.30.Er, 13.25.Ft, 14.40.Lb}

\section{Introduction}	
In the standard model (SM), the mixing of neutral $D$ mesons is due to the fact that their
mass eigenstates ($|D_{1,2}\rangle$)  are not the same as the flavor eigenstates ($\Dz$, $\Dzb$).  
They can be expressed as:
\begin{equation}
|D_{1,2}\rangle = p|\Dz\rangle \pm q |\Dzb\rangle,
\end{equation}
where the complex parameters $p$ and $q$ are obtained from diagonalizing the $\Dz-\Dzb$
mass matrix and $|p|^2+|q|^2=1$ under the assumption of $CPT$ conservation. If \CP\ violation (CPV)
in mixing is neglected, $p$ becomes equal to $q$, so $|D_{1,2}\rangle$ become \CP\ eigenstates,
$\CP|D_{\pm}\rangle=\pm|D_{\pm}\rangle$, and
\begin{equation}
|D_{\pm}\rangle = \frac{1}{\sqrt{2}}[|\Dz\rangle \pm  |\Dzb\rangle].
\end{equation}
The mixing effects can be quantified with two dimensionless parameters $x$ and $y$, defined as:
\begin{equation}
x\equiv \frac{m_1-m_2}{\Gamma},\;\;\; y\equiv\frac{\Gamma_1-\Gamma_2}{2\Gamma},
\end{equation}
where $m_{1,2}$ and $\Gamma_{1,2}$ are the mass and widths of the states $D_{1,2}$ respectively, and
$\Gamma=(\Gamma_1+\Gamma_2)/2$.

In the SM, mixing can occur through short-range box-diagram process, and through long-range
rescattering processes via intermediate hadronic states. The former are highly suppressed by the GIM mechanism 
or by $|V_{ub}V_{cb}|^2$. The latter are difficult to calculate precisely and can be as large as a percent level.
As for the CPV in the charm meson decays, it has been predicted to be rather small in the SM. Significant 
\CP\ violation in the charm meson decays or large mixing parameters 
(Eg: $x\gg 0$ or $y\gg 0$)~\cite{Lenz:2010pr} would be a signature for new physics (NP) beyond the SM.

\section{Measurement of Mixing and CPV in WS $D^0$ Decays}
One typical approach to study mixing and CPV in charm meson decays 
is to examine the decay time distribution of the wrong sign (WS) $\Dz$ decay, such as 
$\Dz\to K^+\pi^-$ from the $D^{*+}$ decay~\cite{charge}.  The WS decay can occur via mixing, where an initially pure 
$\Dz$ oscillates to become a $\Dzb$, then undergoes a Cabibbo-favored (CF) decay
to $K^+\pi^-$; or it can occur via doubly-Cabbibo-suppressed (DCS) decay. 
The $\Dz$ meson is reconstructed in the decay $D^{*+}\to\Dz\pi^+$, the charge of
the soft pion from $D^{*+}$ decay indicates the initial flavor of the $\Dz$ meson. 
Since the mixing rate is small and under the assumption of  no CPV, the time-dependent decay rate 
distribution can be written as:
\begin{equation}
\Gamma(\Dz\to K^+\pi^-)\propto R_D+y^\prime \sqrt{R_D}(\Gamma t)+ \frac{x^{\prime 2}+y^{\prime 2}}{4}(\Gamma t)^2,
\label{eq:decay_rate}
\end{equation}
where $R_D$ is the ratio of DCS and CF decay rates, $x^\prime=x\cos\delta_{D}+y\sin\delta_{D}$,
$x^\prime=-x\sin\delta_{D}+y\cos\delta_{D}$, and $\delta_{D}$ is the strong phase difference
between DCS and CF amplitudes. When allowing for CPV, the form of 
Eq.~\ref{eq:decay_rate} remains the same but has separate coefficients $R^\pm_D$, $x^{\prime 2}_{\pm}$
and $y^{\prime 2}_{\pm}$ for $\Dz$ and $\Dzb$ decays.

Similarly, the mixing parameters of other WS hadronic decays can be also extracted from their
time-dependent rate distribution in a similar fashion. For a multi-body decay such as 
$\Dz\to K^+\pi^-\pi^0$, additional sensitivity can be gained by including the position of
each event within the Dalitz plot in a time-dependent amplitude fit, since the distributions of 
the DCS and CF decays differ. However, one must be careful:  the mixing parameter $x^\prime$ and $y^\prime$ 
measured in $\Dz\to K^+\pi^-\pi^0$ can not be directly compared to the ones in the $\Dz\to K^+\pi^-$ decay,
since their strong phase differences $\delta_D$ are not equal. 
\begin{table}[htb]
\tbl{Summary of the measurements~{\protect \cite{Aubert:2007wf,Aubert:2008zh}}  of mixing and \CP\ violation in $\Dz\to K^+\pi^-$ and
$\Dz\to K^+\pi^-\pi^0$ from \babar .}
{\begin{tabular}{|c|c|c|} \hline
Fit Type & $\Dz\to K^+\pi^-$ [$10^{-3}$] & $\Dz\to K^+\pi^-\pi^0$ [$10^{-2}$] \\\hline\hline
No CPV &  $x^{\prime 2}=-0.22\pm0.30\stat\pm0.21\syst$ & $x^{\prime\prime 2}=2.61^{+0.57}_{-0.68}\stat\pm0.39\syst$ \\\hline
   &  $y^{\prime 2}=9.7\pm4.4\stat\pm3.1\syst$ & $y^{\prime\prime 2}=-0.06^{+0.55}_{-0.64}\stat\pm0.34\syst$ \\\hline\hline
CPV allowed&  $x_{+}^{\prime 2}=-0.24\pm0.43\stat\pm0.30\syst$ & $x^{\prime\prime 2}_+=2.53^{+0.54}_{-0.63}\stat\pm0.39\syst$ \\\hline
           &  $x_{-}^{\prime 2}=-0.20\pm0.41\stat\pm0.29\syst$ & $x^{\prime\prime 2}_+=3.55^{+0.73}_{-0.83}\stat\pm0.65\syst$ \\\hline
           &   $y_{+}^{\prime 2}=9.9\pm6.4\stat\pm4.5\syst$ & $y^{\prime\prime 2}_+=-0.05^{+0.63}_{-0.67}\stat\pm0.50\syst$ \\\hline
           &  $y_{-}^{\prime 2}=9.6\pm6.1\stat\pm4.3\syst$ & $y^{\prime\prime 2}_+=-0.54^{+0.40}_{-1.16}\stat\pm0.41\syst$ \\\hline
\end{tabular} \label{ta1}}\end{table}

Based on a $384\,\invfb$  data sample, the \babar\ experiment performed 
measurements~\cite{Aubert:2007wf,Aubert:2008zh} of mixing parameters and searched for CPV in both 
$\Dz\to K^+\pi^-$ and $\Dz\to K^+\pi^-\pi^0$. The results are summarized
in Table~\ref{ta1}. Both measurements find mixing signals with a significance of 
more than $3\,\sigma$, and see no evidence of CPV.

\section{Measurement of Mixing and CPV in $\Dz\to\KS\pi^+\pi^-,\KS K^+K^-$}
Without the knowledge of the strong phase difference, the mixing parameter $x$ and $y$ can not 
be extracted from the measurements in the decay $\Dz\to K^+\pi^-$ and $\Dz\to K^+\pi^-\pi^0$.
This ambiguity can be resolved by a time-dependent amplitude analysis~\cite{Asner:2005sz} using the 
decay $\Dz\to\KS\pi^+\pi^-$ and $\Dz\to\KS K^+K^-$. Because both the decay final states 
are self-conjugate states that include \CP-even and \CP-odd eigenstates, it allows
the relative phases to be determined. As a result, the mixing parameter $x$ and $y$ can be
measured directly.

With $469\,\invfb$ of data sample, \babar\ performed a measurement~\cite{delAmoSanchez:2010xz} 
of mixing parameters
using the decay  $\Dz\to \KS\pi^+\pi^-$ and $\Dz\to \KS K^+K^-$. We found  that
\begin{equation}
x=(0.16\pm0.23\pm0.12\pm0.08)\%,\;\;\;
y=(0.57\pm0.20\pm0.13\pm0.07)\%,
\end{equation}
where the first error is the statistical uncertainty, the second error is the systematic
uncertainty and the third one is the uncertainty due to the Dalitz models. We also repeated the
fit by allowing for CPV and saw no evidence of CPV. 

\section{Measurement of Mixing in $\Dz$ Lifetime difference}
$D^0$ mixing can be measured by comparing the lifetime extracted from the analysis
of $D^0$ decays into $K^-\pi^+$ and $h^+h^- (h=K,\pi)$ final states. The $K^-\pi^+$ is
a mixed \CP-even and \CP-odd final state, and $h^+h^-$ is a \CP-even final state.
Thus we have 
\begin{equation}
y \approx y_{\CP} \approx \frac{\tau(\Dz\to K^-\pi^+)}
{\tau(\Dz\to h^+h^-)}-1,
\end{equation}
where $\tau$ is the measured $\Dz$ lifetime in the $K^-\pi^+$ and $h^+h^-$ final states.
\babar\ performed a lifetime ratio measurement between $\Dz\to K^-\pi^+$ and
$\Dz\to K^+K^-$ final states using a $384\,\invfb$ data sample~\cite{:2009ck,Aubert:2007en}. We found that
\begin{equation}
y_{\CP}=(1.16\pm 0.22\stat\pm0.18\syst)\%.
\end{equation}
The significance of this result from  no-mixing hypothesis is $4.1\,\sigma$. 

\section{Measurement of Time-Integrated \CP\ Asymmetries}
Another method to search for CPV is to measure the time-integrated \CP\ asymmetry ($A_{\CP}$) of 
$D$ meson decay to a given final state $f$:
\begin{equation}
A_{\CP}=\frac{\Gamma(D\to f)-\Gamma(\bar{D}\to\bar{f})}
{\Gamma(D\to f)+\Gamma(\bar{D}\to\bar{f})},
\end{equation}
where $\Gamma$ is the partial decay width for this decay.
Many searches for time-integrated \CP\ asymmetries have been performed by both \babar\ and 
Belle experiments and no evidence of CPV has been seen yet. So far, all the measurements 
have been limited by the statistical and systematic uncertainties.

Most recently, the \babar\ experiment performed a measurement of time-integrated \CP\ asymmetry
in the decay $D^+\to\KS\pi^+$~\cite{:2010hd}.
In the decay $D^+\to\KS\pi^+$, the SM predicts $A_{\CP}$
to be $(-0.332\pm 0.006)\,\%$, due to CPV in $\Kz-\Kzb$ 
mixing~\cite{Nakamura:2010zzi}. 
However, contributions from non-SM processes may
reduce the value of $A_{\CP}$ or enhance it up to 
the level of one percent~\cite{Bigi:1994aw,Lipkin:1999qz}. 
A significant deviation of the
$A_{\CP}$ measurement from the SM expectation would be evidence
for the presence of NP beyond the SM.
Due to the smallness of the predicted value from the SM,
this measurement requires high statistics and precise
control of the systematic uncertainties. 

We optimize our signal reconstruction efficiency and background rejection 
by using a a Boosted Decision Tree algorithm.
With a $469\,\invfb$ data sample, we reconstruct $(807\pm0.1)\times 10^3$ signal 
events. 

One largest systematic uncertainty in the search for time-integrated
\CP\ asymmetries is the differences in the charged track reconstruction efficiencies.
In this analysis we have developed a data-based method to determine the 
charge asymmetry in track reconstruction as a function of the magnitude of 
the track momentum and its polar angle.
Since $B$ mesons are produced in the process $\epem\to\Y4S\to\BB$ nearly 
at rest in the CM frame and decay isotropically in the $B$ rest frame,
these events provide a high statistics control sample essentially free of 
any physics-induced charge asymmetry.
However, data recorded at the $\Y4S$ resonance also include continuum production
$\epem\to\qqbar\;(q=u,d,s,c)$, where there is a non-negligible forward-backward asymmetry due to 
the interference between the single virtual photon process and other production processes,
as described above. 
The continuum contribution is estimated using the off-resonance data 
rescaled to the same luminosity as the on-resonance data sample.
Subtracting the rescaled off-resonance sample from the on-resonance one, we obtain
the number of reconstructed tracks corresponding to the $B$ meson decays only.
Therefore, the relative detection and identification efficiencies of 
the positively and negatively charged particles for given selection criteria 
can be determined using the numbers of positively and negatively 
reconstructed tracks directly from data.

Using samples, respectively, of $8.5\,\invfb$ on-resonance and 
$9.5\,\invfb$ off-resonance data
and applying the same charged pion track selection criteria used in the reconstruction 
of $D^+\to\KS\pi^+$ decays we obtain a sample of more than 20 million tracks
after the subtraction of the off-resonance sample.
We use this sample to produce a map for the ratio of detection efficiencies
for $\pi^+$ and $\pi^-$ as a function of the track-momentum magnitude 
and $\cos\theta$, where $\theta$ is the polar angle of the track 
in the laboratory frame. With this new method, we were able to control the total systematical
uncertainty in the measurement to be 
less than $0.1\,\%$, and obtained:
\begin{equation}
A_{\CP}(\Dz\to\KS\pi^+)=(-0.44\pm 0.13\stat \pm 0.10\syst)\,\%.
\end{equation}
This measurement is the most precise single measurement of time-integrated
\CP\ asymmetry in charm meson decays so far. The method we developed to measure the 
asymmetries in charged track reconstruction efficiencies can be used as a general method 
in other similar measurements.

\section{Search for \CP\ Violation using $T$-odd Correlations}
Recently the \babar\ experiment performed a search for CPV~\cite{delAmoSanchez:2010xj} by exploring the 
$T$-odd correlation in the decay $\Dz\to K^+K^-\pi^+\pi^-$. We define a kinematic triple 
product correlation $C_T=\vec{p}_{K^+}\cdot (\vec{p}_{\pi+}\times \vec{p}_{\pi^-})$,
where each $\vec{p}_i$ is a momentum vector of one of the particles in the decay. The product is
odd under time-reversal ($T$) with the assumption of CPT invariance, thus 
$T$-violation is a signal of CPV.  A $T$-odd observable is then defined as
\begin{equation}
\mathcal{A}_T=\frac{1}{2}(A_T-\bar{A}_T),
\end{equation}
where $A_T$ and $\bar{A}_T$ are defined as
\begin{equation}
A_T=\frac{\Gamma(C_T>0)-\Gamma(C_T<0)}{\Gamma(C_T>0)+\Gamma(C_T<0)},\;\;
\bar{A}_T=\frac{\Gamma(-\bar{C}_T>0)-\Gamma(-\bar{C}_T<0)}
{\Gamma(-\bar{C}_T>0)+\Gamma(-\bar{C}_T<0)}
\end{equation}
With a $470\,\invfb$ data sample, \babar\ measured that
\begin{equation}
\mathcal{A}_T=(1.0\pm5.1\stat\pm4.4\syst)\times 10^{-3}.
\end{equation}
The result is consistent with the SM expectation.

\section{Conclusion}
Measurement of mixing and CPV in charm meson decays provides new and unique
opportunities to search for NP. In this paper, we give a brief review
of current measurements from the \babar\ experiments. 
These results constrain the possible effects of NP.



\begin{thebibliography}{0}
\bibitem{charge}
 Throughout this paper charge conjugation is implied.
\end{thebibliography}
\end{document}